\documentclass[pre,twocolumn,aps,showpacs]{revtex4-1}
\usepackage{graphicx}
\usepackage{amsmath}
\usepackage{verbatim}
\usepackage{color}
\usepackage[dvipsnames]{xcolor}
\begin{document}
\newcommand{\tca}[1]{\textcolor{blue}{#1}}
\newcommand{\tcom}[1]{{\it\textcolor{magenta}{#1}}}

\title{Nonlinear electrostatic oscillations in a cold magnetized electron-positron plasma}

\author{Prabal Singh Verma\footnote{prabal-singh.verma@univ-amu.fr}}
\affiliation{CNRS/Universit\'e d'Aix-Marseille
Centre saint J\'er\^ome, case 232, F-13397 Marseille cedex 20, France
}
\author{Tapan Chandra Adhyapak}
\affiliation{Institut f\"ur Physik, Johannes Gutenberg-Universit\"at Mainz, Staudingerweg 7-9, 55128 Mainz, Germany}

\date{\today}

\begin{abstract}

We study the spatio-temporal evolution of the nonlinear electrostatic
oscillations in a cold magnetized electron-positron (e-p) plasma using both
analytics and simulations.  Using a perturbative method we demonstrate that the
nonlinear solutions change significantly when a pure electrostatic mode is
excited at the linear level instead of a mixed upper-hybrid and zero-frequency
mode that is considered in a recent study.  The pure electrostatic oscillations
undergo phase mixing nonlinearly. However, the presence of the magnetic field
significantly delays the phase-mixing compared to that observed in the
corresponding unmagnetized plasma.  Using 1D PIC simulations we then analyze
the damping of the primary modes of the pure oscillations in detail and infer
the dependence of the phase-mixing time on the magnetic field and the amplitude
of the oscillations. The results are remarkably different from those found for
the mixed upper-hybrid mode mentioned above. Exploiting the symmetry of the e-p
plasma we then explain a generalized symmetry of our non-linear solutions. The
symmetry allows us to construct a unique nonlinear solution up to the second
order which does not show any signature of phase mixing but results in a
nonlinear wave traveling at upper-hybrid frequency. Our investigations have
relevance for laboratory/astrophysical e-p plasmas.

\end{abstract}

\pacs{52.35.Mw, 52.27.Ny, 52.65.Rr}

\maketitle

\section{Introduction}

The study of nonlinear electrostatic oscillations in a cold plasma is an
interesting area of research from both theoretical and application point of
view \cite{dawson1959nonlinear, davidson1968nonlinear, davidson1972methods, stenflo1998electron, brodin2014large, brodin2014nonlinear, 
brodin2017simple}.
One of the major application of such studies is in the laser (or particle beam)
induced wake-field acceleration techniques where electrostatic oscillations in
plasmas are used to accelerate charged particles to high-energies in short
distances \cite{faure2004laser, caldwell2009proton, nature2014}. 
{However,}
in the above acceleration technique, maximum acceleration can only be achieved
when 
{the}
amplitude of the electrostatic oscillations is kept below a critical value
known as 
{the}
wave-breaking amplitude
\cite{dawson1959nonlinear,davidson1972methods,akhiezer1956theory}.  Beyond the
critical amplitude coherent oscillations break and 
{the}
amplitude of the
accelerating field diminishes dramatically by converting a part of 
{the}
coherent energy into random kinetic energy of the particles
\cite{verma2012residual}. As a result wave is no more able to accelerate
particles to the desired energy.  Physically, beyond the critical amplitude
trajectories of neighboring oscillators taking part in the wave cross each
other, and as a consequence the oscillators loose their coherent motion
\cite{dawson1959nonlinear}.  

There is another physical process however slower but can lead to wave-breaking
of arbitrarily small amplitude oscillations. This process is known as phase
mixing which occurs when the frequency of the oscillations due to any physical
reason (for example -- inhomogeneity in the ion background
\cite{dawson1959nonlinear,kaw1973quasiresonant,infeld1989ion,nappi1991modification},
relativistic mass variation effects
\cite{infeld1989relativistic,sengupta2009phase,verma2012breaking},
inhomogeneous magnetic field \cite{maity2012breaking} etc.) acquires a spatial
dependence. Because of the space-dependent frequency neighbouring oscillators
slowly go out of phase 
{which leads}
to the trajectory crossing and hence
wave-breaking after certain time. Phase mixing is responsible for the flow of
energy irreversibly into higher harmonics and leads to damping of the primary
mode \cite{kaw1973quasiresonant,gupta1999phase,sengupta2009phase}.  Thus, phase
mixing is an undesirable physical process as it can significantly affect the
maximum energy gain in the wake-field acceleration experiments {as suggested in Ref. 
\onlinecite{verma2012breaking}}.   

There is considerable understanding about the phase mixing and nonlinear
evolution of
electrostatic oscillations in the electron-ion (e-i), and other plasmas
{where charged species have unequal masses} \cite{infeld1989ion,
gupta1999phase, nappi1991modification, verma2011nonlinear}. Recent developments
suggest electron-positron (e-p) plasma as an alternative interesting system for
plasma experiments \cite{saberian2013langmuir, helander2014microstability,
edwards2016strongly, wang2016evolution}.  It is possible to produce e-p plasmas
in the laboratory by laser matter interaction \cite{chen2009relativistic}, and
they also exist naturally in various astrophysical environments
\cite{sturrock1971model, michel1982theory, begelman1984theory}.  The dynamics
of the electrostatic oscillations in an e-p plasma differ significantly from
that in the e-i, and other plasmas {referred to above}, because of the
equal mass of the two charged species involved. {Besides, the presence of
a magnetic field offers further richness to the oscillations in an e-p plasma.
An external magnetic field is suggested to play a crucial role in the evolution
of the upper-hybrid mode \cite{maity2012breaking,maity2013wave,karmakar2016wave}, which is a mode
of electrostatic oscillations in a magnetized plasma.} 
Although, upper-hybrid oscillations mixed with a zero frequency mode in a
magnetized e-p plasma have received recent attentions \cite{maity2013phase},
how a {\it pure} upper-hybrid oscillation -- which lacks any other mode and
therefore is easier to excite in experiments -- evolves nonlinearly in space
and time in a magnetized e-p plasma is still not known.

In this paper we investigate the dynamics of a pure upper-hybrid oscillation in
a cold e-p plasma.  Although, electrostatic oscillations in a magnetized e-p
plasma has been studied in Ref.  \onlinecite{maity2013phase}, only a mixed mode
consisting of the upper-hybrid oscillations and a space-dependent zero
frequency DC mode has been examined. For such mixed oscillations, the spatial
dependence of the DC mode present from the beginning is directly responsible
for the wave-breaking through a mechanism mentioned above.  In contrast, here
we choose our initial conditions such that pure upper-hybrid oscillations are
set up at the linear level so that there is no cause of wave-breaking at that
order. We then use a perturbative method to probe how higher order
perturbations affect the pure upper-hybrid mode.

We find that pure upper-hybrid oscillations in a cold e-p plasma phase-mix and
break at an arbitrarily small amplitudes. However,  we demonstrate that, 
{due to}
the presence of the magnetic field, the perturbations to the pure 
{upper-hybrid}
oscillations preserve their oscillatory nature until the second order solution
and hence do not grow in time.  Nevertheless, higher harmonics and mixed modes
are developed as the plasma acquires inhomogeneity due to the imbalance of
ponderomotive forces and nonlinear Lorentz forces. 
{We obtained the analytical
solutions of all the relevant physical quantities up to the third order in a
perturbative analysis. Furthermore, we explain a mathematical symmetry in
the nonlinear upper-hybrid oscillations and demonstrate
it up to the third order.}

{Our analysis of the above results}
imply that the presence of the magnetic field significantly suppresses the
phase-mixing of the pure electrostatic mode  in comparison to the fate of the
electrostatic oscillations in an unmagnetized e-p plasma. For an unmagnetized
e-p plasma, the use of a similar perturbative approach has established that
density fluctuations in both the charged species grow in time as $\sim t^2$
already in the second order solution, indicating a rapid bunching of particles
and hence much rapid phase mixing \cite{stewart1993nonlinear, gupta1999phase,
verma2011nonlinear, maity2014phase, verma2016nonlinear}.  Our results also
reveal significant differences between the evolutions of a pure upper-hybrid
oscillation and a 
{mixed}
upper-hybrid mode accompanied by a space dependent DC term.
In the latter case, 
{the second order solutions grow in time and}
the sum of 
{the}
density fluctuations 
{acquires fast secular terms}
$\sim t$ already
in the third order solutions. 
In contrast, for a pure upper-hybrid mode second order solutions do not
grow in time, and although the third order solutions grow as $t \cos{\omega_h
t}$ where $\omega_h$ is the frequency of the upper-hybrid mode, the sum of density
fluctuations vanishes.

{We then proceeded to}
perform a one dimensional particle-in-cell 
(1D PIC)
{simulation}
\cite{birdsall2004plasma}, 
{first confirming}
a good agreement 
{with our analytical results.}
{Later, we estimate the phase-mixing time and its dependence on the
amplitude of the oscillations and the magnetic field from our simulations.}
Furthermore, 
{using the confirmed symmetry of the oscillations mentioned above, we
identify}
an unique nonlinear solution in a cold magnetized e-p plasma which does not
show any signature of phase mixing but results in a nonlinear wave propagating
at the upper-hybrid frequency.  

{Now we proceed to present our analysis.} 
The flow of the paper is as follows. Section II deals with the linear and
nonlinear solutions of upper-hybrid oscillations in a cold e-p plasma. In
section III a comparison between numerical experiment and analytical results is
provided {and the scalings of the phase-mixing time are estimated}.  
Section IV describes the construction of the nonlinear solution, up
to the second order, which does not show any signature of phase mixing. Section
V contains the summary 
of the results obtained in this work.     

\section{Governing equations and perturbation analysis}
The basic equations describing the dynamics of a cold magnetized e-p plasma are
the continuity equations,   
\begin{equation}\label{econ}
 {\partial_t n_{e}}+\boldsymbol{\nabla}(n_{e}{\bf v_{e}})=0,
\end{equation}
\begin{equation}\label{icon}
 {\partial_t n_{p}}+\boldsymbol{\nabla}(n_{p}{\bf v_{p}})=0,
\end{equation}
{the}
momentum equations, 
\begin{equation}\label{emom}
({\partial_t } + {\bf v_e}.\boldsymbol{\nabla}){\bf v_e}= -\frac{e}{m} [{\bf E} + (1/c){\bf v_e}\times {\bf B}],
\end{equation}
\begin{equation}\label{imom}
({\partial_t } + {\bf v_p}.\boldsymbol{\nabla}){\bf v_p}= \frac{e}{m} [{\bf E} + (1/c){\bf v_p}\times {\bf B}],
\end{equation}
and the Poisson's equation,
 \begin{equation}\label{poi}
   \boldsymbol{\nabla}.{\bf E}= 4 \pi e (n_p - n_e).
 \end{equation}
Here the subscript `$p$', stands for the positron and `$e$', stands for the
electron.  The densities and velocities of both the species are denoted by
$n_\alpha$ and $\mathbf{v_\alpha}$ respectively, where $\alpha$ is the
corresponding subscript.  The quantity $c$ is the speed of light in vacuum,
{$e$ is the charge of a positron (not to be confused with the subscript
`$e$')}, ${\bf E}$ the electric field and ${\bf B}$ is a homogeneous external
magnetic field applied along the $z$-direction i.e. ${\bf B} = B_0 \hat z$.
{\it CGS} unit is used throughout. 

Since we are interested only in the electrostatic mode, spatial variations are
restricted to one direction, which we consider to be along the $x$-axis,
without any loss of generality.  Thus, the set of equations
\eqref{econ}-\eqref{poi} reduces to the following equations,
\begin{equation}\label{econ1}
 {\partial_t n_{e}}+{\partial_x(n_{e}v_{ex})}=0,
\end{equation}
\begin{equation}\label{icon1}
 {\partial_t n_{p}}+{\partial_x(n_{p}v_{px})}=0,
\end{equation}
\begin{equation}\label{emom1_x}
{\partial_t v_{ex}} + v_{ex}{\partial_x v_{ex}}= -\frac{e}{m} E_x - \omega_c v_{ey},  
\end{equation}
\begin{equation}\label{imom1_x}
{\partial_t v_{px}} + v_{px}{\partial_x v_{px}}=  \frac{e}{m} E_x + \omega_c v_{py}, 
\end{equation}
\begin{equation}\label{emom1_y}
{\partial_t v_{ey}} + v_{ex}{\partial_x v_{ey}}=   \omega_c v_{ex},  
\end{equation}
\begin{equation}\label{imom1_y}
{\partial_t v_{py}} + v_{px}{\partial_x v_{py}}=  - \omega_c v_{px}, 
\end{equation}
 \begin{equation}\label{poi1}
  {\partial_x E_x}= 4 \pi e (n_p - n_e),
\end{equation}
where $\omega_c = e B_0/mc$ is the cyclotron frequency, with $m$ the common
mass of both an electron and a positron.  Now we proceed to adopt the
perturbative analysis in order to obtain nonlinear solutions for the set of
equation \eqref{econ1}-\eqref{poi1}. In such analysis the general solutions can
be expressed as, 
\begin{equation}\label{sol1}
n_{e}(x,t) = n_0 + n_{e}^{(1)}(x,t) + n_{e}^{(2)}(x,t) + n_{e}^{(3)}(x,t) + ...
 \end{equation}
 \begin{equation}\label{sol2}
n_{p}(x,t) = n_0 + n_{p}^{(1)}(x,t) + n_{p}^{(2)}(x,t) + n_{p}^{(3)}(x,t) + ...
\end{equation}
Here $n_0$ is the equilibrium density of both the species and the superscripts
{$(\beta)$} denote the order of the solutions. Different orders of the
solutions are understood to be proportional to the corresponding powers of a
small parameter, $\delta$, which 
{is}
introduced below. 
Solutions for other physical quantities can similarly
be written down.  We now introduce the variable $n_{d}$, which is proportional
to the net charge density, such that, 
\begin{equation}
n_{d}(x,t) = n_{p}(x,t)-n_{e}(x,t). \nonumber 
\end{equation}
Since this variable will later be used in the analysis we explicitly write down
its general solution as, 
\begin{equation}\label{sol3}
n_{d}(x,t) = n_{d}^{(1)}(x,t) + n_{d}^{(2)}(x,t) + n_{d}^{(3)}(x,t) + ...
\end{equation}

\subsection{First order solution}
\label{sec:1storder}

The set of equations \eqref{econ1}-\eqref{poi1} in the first order
approximation can be expressed as, 
\begin{equation}\label{econ2}
 {\partial_t n_{e}^{(1)}}+n_{0}{\partial_x v_{ex}^{(1)}}=0,
\end{equation}
\begin{equation}\label{icon2}
 {\partial_t n_{p}^{(1)}}+n_{0}{\partial_x v_{px}^{(1)}}=0,
\end{equation}
\begin{equation}\label{emom2_x}
{\partial_t v_{ex}^{(1)}} = -\frac{e}{m} E_x^{(1)} - \omega_c v_{ey}^{(1)},  
\end{equation}
\begin{equation}\label{imom2_x}
{\partial_t v_{px}^{(1)}} =  \frac{e}{m} E_x^{(1)} + \omega_c v_{py}^{(1)}, 
\end{equation}
\begin{equation}\label{emom2_y}
{\partial_t v_{ey}^{(1)}} =   \omega_c v_{ex}^{(1)},  
\end{equation}
\begin{equation}\label{imom2_y}
{\partial_t v_{py}^{(1)}} =  - \omega_c v_{px}^{(1)}, 
\end{equation}
 \begin{equation}\label{poi2}
  {\partial_x E_x^{(1)}} = 4 \pi e (n_p^{(1)} - n_e^{(1)}) = 4 \pi e n_d^{(1)}.
\end{equation}
Equations \eqref{econ2}-\eqref{imom2_x} and \eqref{poi2} are combined to get,
\begin{equation}\label{pdend1}
 \partial_t^2 \delta n_{d}^{(1)} + 2 \omega_{p}^2\delta n_{d}^{(1)} + \omega_c n_{0} \partial_x (v_{ey}^{(1)}+v_{py}^{(1)}) =0,
\end{equation}
where,  $\omega_{p} = \sqrt{4 \pi n_{0} e^2/{m}}$ is the plasma frequency of
either of the species.  Now equations \eqref{emom2_y}-\eqref{imom2_y} and
equations \eqref{econ2}-\eqref{icon2} give,
\begin{equation}\label{pdend2}
 \partial_t (n_0 \partial_x (v_{ey}^{(1)} + v_{py}^{(1)}) = \partial_t(\omega_c n_{d}^{(1)}) ,
\end{equation}
Integrating equation \eqref{pdend2} w.r.t. $t$ we get,
\begin{equation}\label{pdend3}
 n_0 \partial_x (v_{ey}^{(1)} + v_{py}^{(1)}) = \omega_c n_{d}^{(1)} + C_1.
\end{equation}
Here the constant $C_1$ has to be determined from the initial conditions.  In
order to see pure upper-hybrid oscillations in the linear solution, initial
conditions need to be chosen such that $C_1$ becomes zero.  Otherwise, a mixed
mode results in
DC terms which can 
trigger phase mixing nonlinearly \cite{maity2013phase}.  Therefore, we choose
to perturb the system as follows, 
\begin{eqnarray}
 \nonumber n_e(x,0) = n_{0},v_{ex}(x,0) = \frac{\delta}{2}\frac{\omega_h}{k} \sin(kx), v_{ey}(x,0) = 0, \\
 \nonumber n_p(x,0) = n_{0}, v_{px}(x,0) = -\frac{\delta}{2}\frac{\omega_h}{k}  \sin(kx),v_{py}(x,0) = 0,   
\end{eqnarray}
where  $\omega_h = \sqrt{(2 \omega_p^2 + \omega_c^2)}$ and $\delta$ is {a
small parameter controlling the amplitudes of the perturbations.} 
For the chosen initial conditions equation \eqref{pdend3} reduces to,
\begin{equation}\label{pdend4}
 n_0 \partial_x (v_{ey}^{(1)} + v_{py}^{(1)}) = \omega_c n_{d}^{(1)}.
\end{equation}
From equations \eqref{pdend1} and \eqref{pdend4} we get, 
\begin{equation}\label{pdend1}
 \partial_t^2  n_{d}^{(1)} + \omega_{h}^2 n_{d}^{(1)} =0. 
\end{equation}
Thus, the quantity $\omega_h$, defined earlier, is the upper-hybrid frequency.
Now the solution in the first order becomes, 
\begin{eqnarray}
\label{solne11}
  n_{e}^{(1)} &=& -\frac{n_{0}\delta}{2} \cos(kx) \sin(\omega_{h}t), \\
\label{solnp11}
   n_{p}^{(1)} &=& \frac{n_{0}\delta}{2} \cos(kx) \sin(\omega_{h}t), \\
\label{solvex11}
   v_{ex}^{(1)} &=& \frac{\delta}{2}\frac{\omega_h}{k} \sin(kx) \cos(\omega_{h}t), \\
\label{solnvpx11}
   v_{px}^{(1)} &=& -\frac{\delta}{2}\frac{\omega_h}{k} \sin(kx) \cos(\omega_{h}t), \\
\label{solvey11}
   v_{ey}^{(1)} &=& \frac{\delta}{2}\frac{\omega_c}{k} \sin(kx) \sin(\omega_{h}t), \\
\label{sollvpy11}
   v_{py}^{(1)} &=& \frac{\delta}{2}\frac{\omega_c}{k} \sin(kx) \sin(\omega_{h}t), \\
\label{solE11}
  E_x^{(1)} &=& \frac{4\pi e n_{0} \delta}{k} \sin(kx) \sin(\omega_h t).
\end{eqnarray}
From the linear solution we see that all the physical quantities oscillate
coherently with the upper-hybrid frequency and there is no cause of phase
mixing at the linear level.  In the next subsection second order solution is
obtained.

\subsection{Second order solution}
\label{sec:2ndorder}

The set of equations \eqref{econ1}-\eqref{poi1} in the second order
approximation can be expressed as,
\begin{equation}\label{econ4}
 {\partial_t n_{e}^{(2)}}+n_{0}{\partial_x v_{ex}^{(2)}} + \partial_x ( n_{e}^{(1)}v_{ex}^{(1)}) =0,
\end{equation}
\begin{equation}\label{icon4}
 {\partial_t n_{p}^{(2)}}+n_{0}{\partial_x v_{px}^{(2)}}  + \partial_x ( n_{p}^{(1)}v_{px}^{(1)}) =0,
\end{equation}
\begin{equation}\label{emom4_x}
{\partial_t v_{ex}^{(2)}} + v_{ex}^{(1)}{\partial_x v_{ex}^{(1)}}= -\frac{e}{m} E_x^{(2)} - \omega_c v_{ey}^{(2)},  
\end{equation}
\begin{equation}\label{imom4_x}
{\partial_t v_{px}^{(2)}} + v_{px}^{(1)}{\partial_x v_{px}^{(1)}}=  \frac{e}{m} E_x^{(2)} + \omega_c v_{py}^{(2)}, 
\end{equation}
\begin{equation}\label{emom4_y}
{\partial_t v_{ey}^{(2)}} + v_{ex}^{(1)}{\partial_x v_{ey}^{(1)}}=   \omega_c v_{ex}^{(2)},  
\end{equation}
\begin{equation}\label{imom4_y}
{\partial_t v_{py}^{(2)}} + v_{px}^{(1)}{\partial_x v_{py}^{(1)}}=  - \omega_c v_{px}^{(2)}, 
\end{equation}
 \begin{equation}\label{poi4}
  {\partial_x E_x^{(2)}}= 4 \pi e (n_p^{(2)} - n_e^{(2)}) = 4 \pi e n_d^{(2)} ,
 \end{equation}
From the above equations one can easily deduce that, 
\begin{eqnarray}\label{nd2}
n_{d}^{(2)} = 0,\hspace{0.5cm}   
i.e. \hspace{0.5cm} n_{p}^{(2)} = n_{e}^{(2)}. 
\end{eqnarray}
Similar results have also been observed in the absence of a magnetic field, for
example by substituting $\Delta = 1$ in Ref. \onlinecite{verma2011nonlinear}.
Now using equation \eqref{nd2} back in equations \eqref{econ4}-\eqref{poi4} we
obtain, 
\begin{eqnarray}\label{E2}
E_x^{(2)} = 0,  
\hspace{0.5cm} v_{ex}^{(2)} =  v_{px}^{(2)}, 
\hspace{0.5cm} v_{ey}^{(2)} = - v_{py}^{(2)}. 
\end{eqnarray}
So, we observe a symmetry in the nonlinear solution for the pure upper-hybrid
oscillations set up in the first order solution.  Moreover, since the nonlinear
electric field $E_x^{(2)}$, which is responsible for the plasma frequency
contribution in the second order solution,  vanishes self-consistently ,
upper-hybrid mode reduces to cyclotron mode in the dynamical equation.  
Now from equations
\eqref{econ4}-\eqref{E2} we write down the second order solution as follows,
\begin{eqnarray}
\label{solne22}
n_{e}^{(2)} &=& \frac{n_{0}\delta^2 \omega_{h}^2}{\omega_{c}^2-4\omega_{h}^2} \cos(2kx) 
                 \left[\frac{\omega_{p}^2}{\omega_{c}^2} \cos(\omega_{c}t)\right. \nonumber \\ 
             &&  \left. + \frac{3}{8}\cos(2\omega_{h}t)\right] 
                 + \frac{n_{0}\delta^2 \omega_{h}^2}{8 \omega_{c}^2} \cos(2kx),  \\  
\label{solvex22}
v_{ex}^{(2)} &=& \frac{\delta^2 \omega_{h}^2\omega_{p}^2}{2 k \omega_{c}(\omega_{c}^2-4\omega_{h}^2)} \sin(2kx) 
                  \sin(\omega_{c}t) \nonumber \\
             &&  +  \frac{\delta^2 \omega_{h}(\omega_{c}^2+2\omega_{h}^2)} {16 k (\omega_{c}^2-4\omega_{h}^2)} \sin(2kx) 
                    \sin(2 \omega_{h}t), \\
\label{solvey22}
v_{ey}^{(2)} &=& -\frac{\delta^2 \omega_{c} \omega_{h}^2}{2k(\omega_{c}^2-4\omega_{h}^2)} \sin(2kx) 
                  \left[\frac{\omega_{p}^2}{\omega_{c}^2} \cos(\omega_{c}t) \right. \nonumber \\  
             &&  + \left. \frac{3}{8}\cos(2\omega_{h}t)\right] 
                 - \frac{\delta^2 \omega_{h}^2}{16 k \omega_{c}} \sin(2kx).   
\end{eqnarray}
Eqs.  \eqref{nd2}-\eqref{solvey22} provide the second order solution for all
the physical quantities.

{We}
observe that the pure upper-hybrid oscillations in the linear level develops
into a mixed mode  due to the presence of 
{the}
slow 
DC terms and the cyclotron mode $\omega_{c}$.  We also observe
the generation of 
{the}
second harmonics in space and in the upper-hybrid frequency $\omega_h$.
Remarkably, however, there is no time dependent amplitudes in the second order
solutions and hence the perturbations to the pure upper-hybrid oscillations are
oscillatory in nature and do not grow in time up to the second order.
This is in contrast to the non-linear evolution of the mixed upper-hybrid
oscillations studied in Ref. \onlinecite{maity2013phase}.  
In the mixed mode study, the zero frequency mode interacts nonlinearly with
upper-hybrid mode and leads to the generation of terms growing in time $\sim t
\sin(\omega_h t)$ and $\sim t \cos(\omega_h t)$ in the second order solutions.

We note that the nonlinear generation of a DC term in the second order solution
can also be seen in the case of upper-hybrid oscillations in a cold
electron-ion plasma when ions are considered to be infinitely massive and
external magnetic field is homogeneous. {This can be seen from a Taylor expansion of 
the exact 
nonlinear solution considering initial amplitude to be small \cite{davidson1968nonlinear}}.  However, the DC term
gets canceled self-consistently after one complete period, therefore frequency
of the system does not acquire any spatial dependence. As a result, coherent
oscillations are maintained indefinitely at the upper-hybrid frequency in such
plasmas.  On the contrary, in the present study the nonlinear generation of the
cyclotron mode $\omega_{c}$ which is not a harmonic of upper-hybrid mode
$\omega_{h}$ allows the plasma to acquire space dependent frequency due to the
presence of two incommensurate time scales.  Therefore, nonlinear generation of
the cyclotron mode in the second order solution is responsible for the two
simultaneous effects. It is opposing the generation of secular terms $(\sim
t)$, $(\sim t^2)$, as explained below, but making the frequency of the system
space-dependent. 
Thus, though we initiate pure upper-hybrid oscillations in a cold homogeneous
e-p plasma, it nonlinearly acquires inhomogeneity in space which in turn
introduces a spatial dependency in the frequency.  This suggests that
upper-hybrid oscillations in an e-p plasma will always phase mix and break at
arbitrarily small amplitude.     

{The absence of the time dependent amplitudes and the nonlinear generation
of the DC terms in the second order solutions can be understood as follows.}
Since, $E_x^{(2)} =0$, 
{when the magnetic field also zero}
there is nothing on the r.h.s. of \eqref{emom4_x} to balance the 
{effect of the}
ponderomotive forces, represented by $v_{px}^{(1)}{\partial_x v_{px}^{(1)}}$.
As a result $ v_{ex}^{(2)}$ acquires a secular term $\sim t$ which in turn
introduces faster terms $\sim t^2$ in the electron density in equation
\eqref{econ4} \cite{verma2011nonlinear,verma2016nonlinear}.  However, in the
presence of 
{a}
magnetic field 
{Eq. \eqref{emom4_x} is coupled to the Eq. \eqref{solvey22} through
$v_{py}^{(2)}$. In that case, the}  
slow component of the
Lorentz force [see equation \eqref{solvey22}] balances the ponderomotive forces
and the fast component allows the plasma species to oscillate at cyclotron
frequency $\omega_{c}$.  As a result a mixture of cyclotron and upper-hybrid
modes appears in equation \eqref{solvex22} but no term that grows in time is
generated.   
Next, the DC term in \eqref{solvey22} is originated through an indirect
effect of the ponderomotive forces {which becomes clear when one writes 
down a second order dynamical equation for $v_{ey}^{(2)}$. The DC term in
$v_{ey}^{(2)}$ in turn}
makes the density inhomogeneous nonlinearly as can be seen from the following
relation obtained from combining Eqs. \eqref{solne22} and \eqref{solvey22}:
\begin{equation}\label{imom4_y_ne}
{\partial_x v_{ey}^{(2)}} =  - \omega_c n_{e}^{(2)}. 
\end{equation}
Thus, the indirect effect of the ponderomotive forces and self-consistently
generated cyclotron mode make the plasma inhomogeneous nonlinearly in the
second order solution. 
Physically, ponderomotive force is trying to accelerate the charged particles
so as to increase their bunching in time, however the Lorentz force opposes the
acceleration by changing the trajectory of the charged particles. As a result
particle bunching gets slower in the magnetized plasma.

{We}
note that 
{by taking}
the limit $\omega_{c} \rightarrow 0$ in
Eqs. \eqref{solvex22} and  \eqref{solne22} 
{we can recover the}
secular terms ($\sim t$ and
$\sim t^2$ respectively) 
in both the equations.  Thus, 
{our}
results are consistent with 
{the}
previous studies
\cite{verma2011nonlinear,verma2016nonlinear}.  In the next subsection we
proceed to obtain the third order solution 
{and provide an explanation deciphering the observed symmetry in the
solutions.}
 
\subsection{Third order solution}
\label{sec:3rdorder}

The set of equations \eqref{econ1}-\eqref{poi1} in the third order
approximation can be expressed as,
\begin{equation}\label{econ5}
 {\partial_t n_{e}^{(3)}}+n_{0}{\partial_x v_{ex}^{(3)}} + \partial_x ( n_{e}^{(1)}v_{ex}^{(2)} + n_{e}^{(2)}v_{ex}^{(1)}) =0,
\end{equation}
\begin{equation}\label{icon5}
 {\partial_t n_{p}^{(3)}}+n_{0}{\partial_x v_{px}^{(3)}} + \partial_x ( n_{p}^{(1)}v_{px}^{(2)} + n_{p}^{(2)}v_{px}^{(1)}) =0,
\end{equation}
\begin{equation}\label{emom5_x}
{\partial_t v_{ex}^{(2)}} + {\partial_x (v_{ex}^{(1)}v_{ex}^{(2)})}= -\frac{e}{m} E_x^{(3)} - \omega_c v_{ey}^{(3)},  
\end{equation}
\begin{equation}\label{imom5_x}
{\partial_t v_{px}^{(2)}} + {\partial_x (v_{px}^{(1)}v_{px}^{(2)})}= \frac{e}{m} E_x^{(3)} + \omega_c v_{py}^{(3)},  
\end{equation}
\begin{equation}\label{emom5_y}
{\partial_t v_{ey}^{(3)}} + v_{ex}^{(1)}{\partial_x v_{ey}^{(2)}} + v_{ex}^{(2)}{\partial_x v_{ey}^{(1)}}=   \omega_c v_{ex}^{(3)},  
\end{equation}
\begin{equation}\label{imom5_y}
{\partial_t v_{py}^{(3)}} + v_{px}^{(1)}{\partial_x v_{py}^{(2)}} + v_{px}^{(2)}{\partial_x v_{py}^{(1)}}=  -\omega_c v_{px}^{(3)},  
\end{equation}
 \begin{equation}\label{poi5}
  {\partial_x E_x^{(3)}}= 4 \pi e (n_p^{(3)} - n_e^{(3)}) = 4 \pi e n_d^{(3)} ,
 \end{equation}
From the above equations we obtain the following symmetry in the third order
solutions: 
\begin{eqnarray}\label{E5}
v_{ex}^{(3)} =  - v_{px}^{(3)}, 
\hspace{0.5cm} v_{ey}^{(3)} =  v_{py}^{(3)}, 
\hspace{0.5cm} n_{e}^{(3)} =  -n_{p}^{(3)}. 
\end{eqnarray}
{The above symmetry is observed to be complementary to that of the second
order solutions [Eqs. \eqref{nd2} and \eqref{E2}]. The} 
symmetry 
can be extended to any order as:
\begin{eqnarray}
\label{symmetry1}
n_e^{(r)}    &=& (-1)^{r} n_p^{(r)}, \hspace{0.5cm} v_{ex}^{(r)} = (-1)^{r}  v_{px}^{(r)}, \nonumber \\
v_{ey}^{(r)} &=& (-1)^{r+1} v_{py}^{(r)}, 
\end{eqnarray}
where `$r$' is the order of the solution. The symmetry in the solutions results
from the underlying symmetry of the dynamical Eqs. \eqref{econ1}-\eqref{poi1}.
{The symmetry of the dynamical equations in turn is related to the similarity}
of an electron and a positron which differ only in
the sign of their charges. Thus, at any instant if we replace each electron
with a positron and vice versa, $n_e \to n_p^f$ and $n_p \to n_e^f$, where the
subscript `$f$' refers to the values obtained after the operation. But in this
case the form of the equations are preserved only if $v_{ex} \to v_{px}^f$,
$v_{px} \to v_{ex}^f$, $v_{ey} \to - v_{py}^f$, $v_{py} \to - v_{ey}^f$. Thus
the quantities, $(n_p + n_e), (v_{px} + v_{ex})$ and $(v_{py} - v_{ey})$
preserve their values under the operation, 
{
\begin{eqnarray}
n_p    + n_e       &=& n_p^f    + n_e^f, \nonumber \\
v_{px} + v_{ex}    &=& v_{px}^f + v_{ex}^f, \\
v_{py} - v_{ey}    &=& v_{py}^f - v_{ey}^f, \nonumber
\end{eqnarray}}
whereas the quantities, $(n_p - n_e), (v_{px} - v_{ex})$ and $(v_{py} +
v_{ey})$ change their signs,
{
\begin{eqnarray}
n_p    - n_e       &=& -(n_p^f    - n_e^f), \nonumber \\
v_{px} - v_{ex}    &=& -(v_{px}^f - v_{ex}^f), \\
v_{py} + v_{ey}    &=& -(v_{py}^f + v_{ey}^f). \nonumber
\end{eqnarray}}
From the initial conditions used, we see that $\delta \to -\delta$ under the
exchanged operation performed above. Thus, in an expansion in powers of
$\delta$ the quantities $(n_p + n_e), (v_{px} + v_{ex})$ and $(v_{py} -
v_{ey})$ should have only even powers of $\delta$ and the quantities, $(n_p -
n_e), (v_{px} - v_{ex})$ and $(v_{py} + v_{ey})$ possess only odd powers. The
symmetries in Eq.  \eqref{symmetry1} then follow immediately from the
perturbation expansion used for each quantity (see, for example,  Eqs.
\eqref{sol1} and \eqref{sol2}), which are 
expansions in powers of the small parameter $\delta$ of our problem.

{Notably, an immediate consequence of the above symmetry is that the}
sum of the density fluctuations
$(n_{e}^{(3)} + n_{p}^{(3)})$ in the third order vanishes self-consistently.
This is in contrast to the
situation of the mixed mode studied in Ref. \onlinecite{maity2013phase} where
$(n_{e}^{(3)} + n_{p}^{(3)})$ {is found to be $\propto t$}. 
The third order solutions 
{obtained by us}
have rather cumbersome forms. 
To facilitate our discussion below we provide the
expressions for $n_d^{(3)}$ and $E_x^{(3)}$ in the Appendix.
{Note that although the third order solutions contains terms proportional
to $t\cos{\omega_h t}$ growing in time, there is no fast secular term(s)
proportional to $\sim t$ or $t^2$ until the third order. This makes a proper
estimation \cite{gupta1999phase}, of the phase mixing time analytically
difficult.  Therefore,}
we choose to estimate the scaling of phase mixing numerically using 
1D PIC simulation. In the next section, we first make a comparison between 
{the}
analytical results and 
{our} 
PIC simulation. 
{We then proceed to investigate the evolution of the pure upper-hybrid oscillations and } 
estimate the phase mixing time 
{from our}
simulations.  

\section{1D PIC Simulation}

\begin{figure}

\includegraphics[width=0.7\columnwidth]{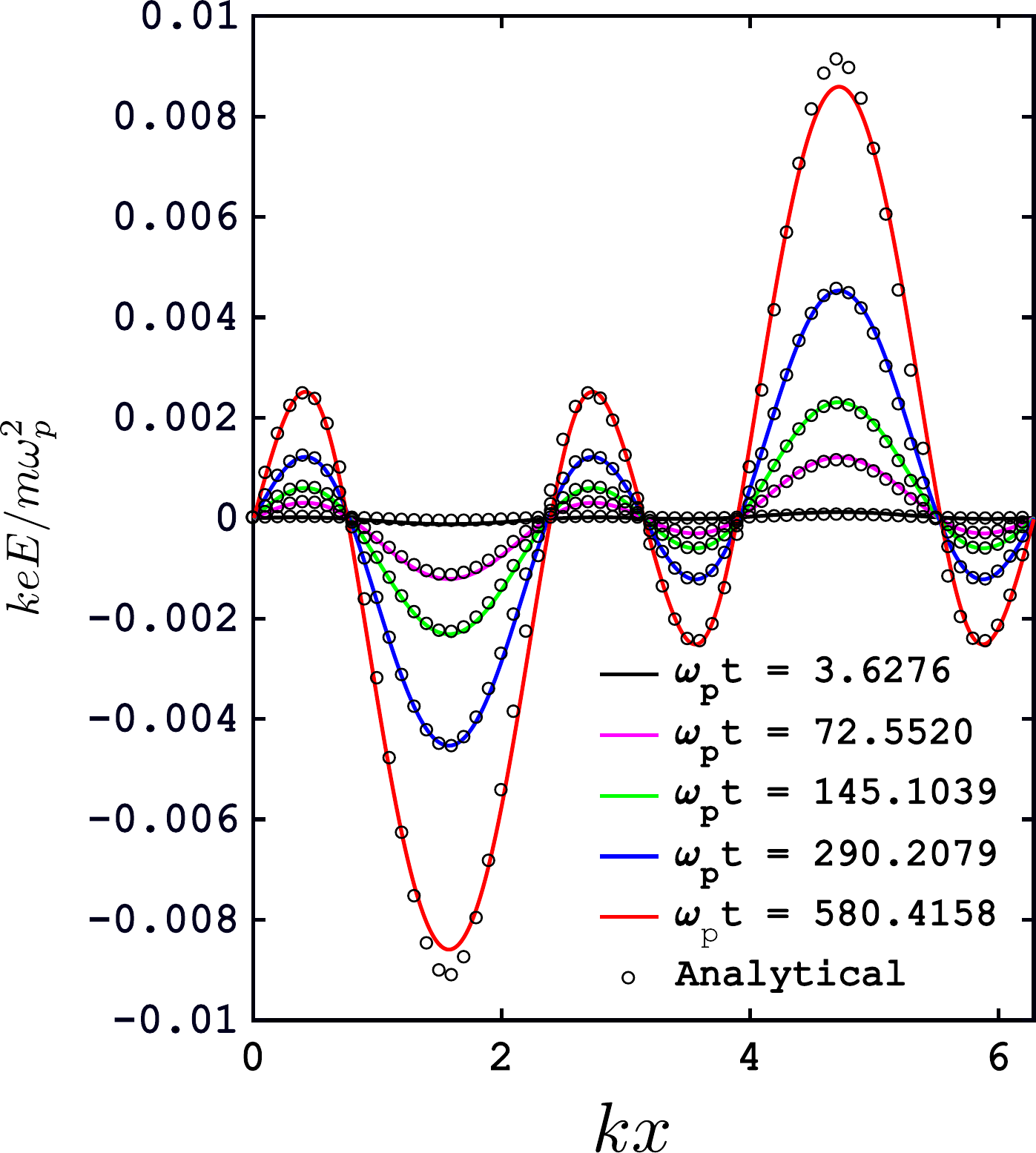}

\caption{Comparison between the numerical and analytical results. The normalized
electric field $k e E/(m\omega_p^2)$ plotted as a function of the normalized
position $kx$ where $k$ is the wave number of the longest (fundamental) mode.
Solid lines are obtained from the simulation and the empty circles represent
the analytical results. Different plots correspond to various time $t$ which
are integral multiple of the upper-hybrid period.}

\label{fig:figure1}

\end{figure}

\begin{figure*}

\includegraphics[width=2\columnwidth]{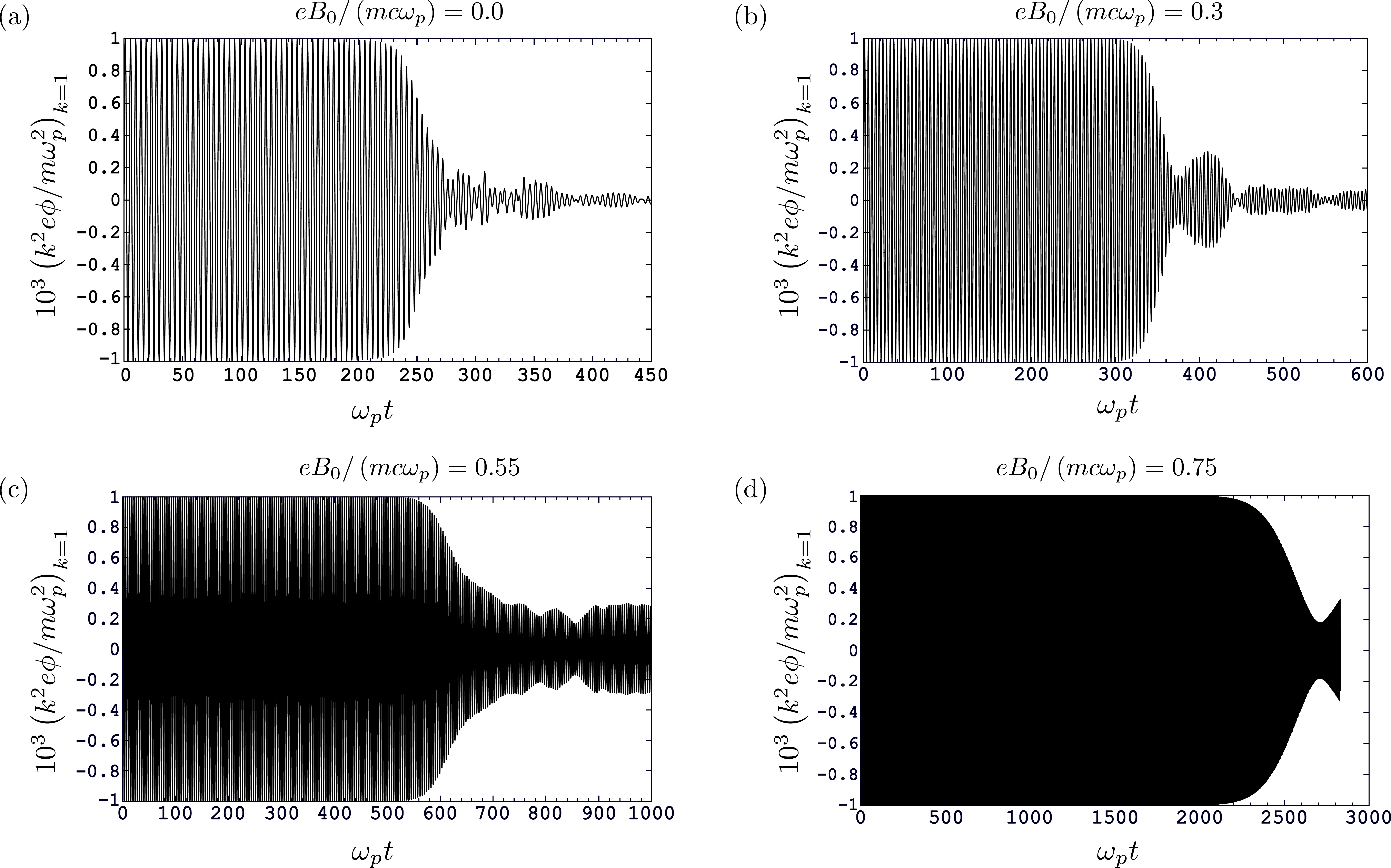}

\caption{ {Time evolution of the primary mode $(k = 1)$ of the normalized
electrostatic potential $k^2 e \phi/m\omega_p^2$  with the normalized external
magnetic field $eB_0/(mc\omega_p)$  set equal to (a) 0.0, (b) 0.3, (c) 0.55 and
(d) 0.75. In all the plots $ \delta = 0.001$.}}

\label{fig:figure2}

\end{figure*}

In order to provide a comparison with our analytical results  we carry out 1-D 
{particle-in-cell (PIC)} 
simulation \cite{birdsall2004plasma}
with periodic boundary conditions. 
{We start our simulations with}
the same initial conditions
{that we used for our analysis above}.
Our simulation parameters are as follows: total number of both kind of
particles 
{N} 
$\sim$ 4$\times10^4$, number of grid points 
{NG} 
$\sim$ 4$\times10^3$, time step $\Delta t$ $\sim$ $\pi/50$. Normalization is as
follows.  $ x \rightarrow kx$, $t \rightarrow \omega_{p}t$, $n_e \rightarrow
n_e/n_0$, $v_e \rightarrow v_e/(\omega_{p}k^{-1})$ and $E \rightarrow
keE/(m\omega_{p}^2)$, where $\omega_{p}$ is the plasma frequency of either of
the species and $k$ is the wave number of the longest (fundamental) mode.

In the first numerical experiment maximum amplitude of the electric field
$keE/(m\omega_{p}^2) = \delta$ and amplitude of the external magnetic field
$eB_0/(m c \omega_{p}) = \omega_c/\omega_{p}$ are chosen to be $0.04$ and $1$,
respectively.  In Fig. \ref{fig:figure1} we show the space-time evolution of
the electric field over several upper-hybrid periods where solid lines are the
results from the PIC simulation and circles are the results from the nonlinear
perturbative analysis up to the third order.  Here, we choose to compare the
results  at times which are integral multiple of the upper-hybrid period $T_h$
such that the contribution of the first order electric-field vanishes.  On the
other hand, the second order electric field is always zero as we have shown in
the second order solutions.  Thus, it is the third order electric field
$E_x^{(3)}$ which is playing the main role in Fig. \ref{fig:figure1}.
{We find}
that the amplitude of the
electric field, which should remain zero in the case of 
{the}
pure oscillations, 
{instead increases}
with time. 
{Furthermore, we clearly see that there is a good agreement between the
analytical and numerical results} 
up to $80$ upper-hybrid periods. 
{However, the results start to deviate from $160$ upper-hybrid periods,
indicating that the effect of the higher order solutions begin to be
significant at this point and beyond.}

{The above observation also indicates}
that the energy is flowing irreversibly into higher harmonics thereby damping
the primary mode -- a signature of phase mixing \cite{kaw1973quasiresonant}.
Since it is difficult to follow the trajectory crossing
\cite{dawson1959nonlinear}, we define 
{the}
phase mixing time as the time when
{the}
amplitude of the primary mode in any physical quantity drops below 
$1/e$
of its initial value.  
To investigate the damping and phase-mixing 
{for varying $\delta$ and $eB_0/(m c \omega_{p})$,}
we observe the time evolution
of the primary mode $(k = 1)$ of the electrostatic potential, $\phi$,
{ which is measured in our simulations as a primary quantity from which the
electric field is derived.}
In Fig. \ref{fig:figure2} we
present four typical time evolution of the primary mode of $\phi$ corresponding
to $eB_0/(m c \omega_{p}) = 0, 0.3, 0.55$ and $0.75$ respectively, for a fixed
$\delta=0.001$. The plots show clear evidence of the damping of the primary
mode after a certain time that varies for each of these cases.

{We can understand the physical reason for the damping of the primary mode
triggered by the phase-mixing as follows. In the absence of an external
magnetic field ponderomotive forces create bunching of the charged particles
leading to inhomogeneity and hence phase-mixing very quickly [see
Fig\ref{fig:figure2} (a)]. However, when we increase $B_0$ while keeping the
maximum amplitude of the electric field $\delta$ constant, we actually increase
the Lorentz force but keep the ponderomotive force $\sim \delta^2$ same. The
Lorentz force opposes the ponderomotive force. Therefore, as we increase $B_0$
the bunching of the particles is increasingly weakened  and hence 
{the process of} 
phase-mixing gets progressively delayed. We can infer the resultant increase in
the phase-mixing time as a function of $B_0$ from Fig. \ref{fig:figure2} where
the damping of the primary mode of $\phi$ is observed to be progressively
delayed from panels (a) {to} (d).} 

\begin{figure}

\includegraphics[width=1\columnwidth]{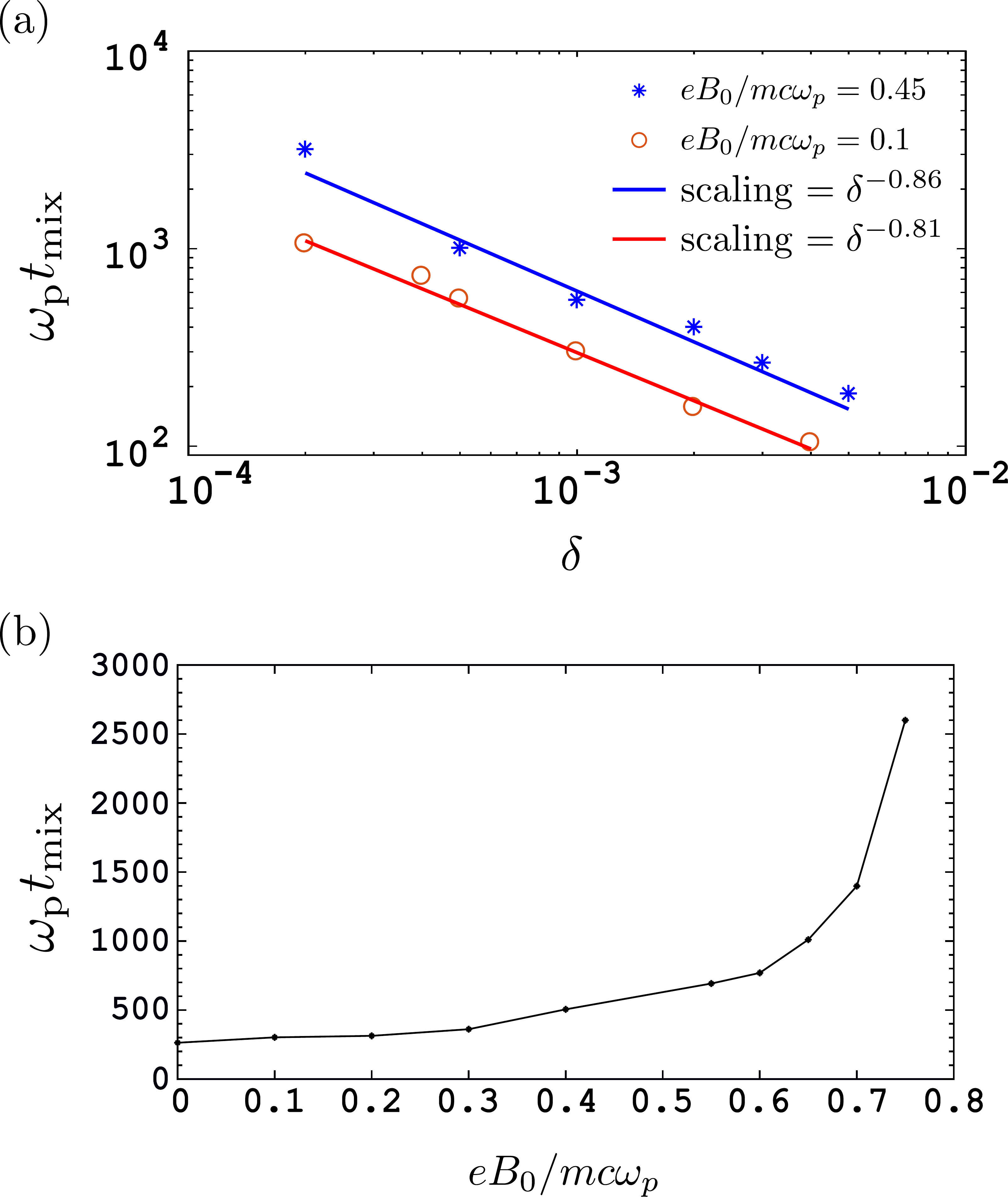}

\caption{(a) Logarithmic plots of the phase-mixing time versus the
maximum amplitude of the electric field $\delta$. The red (`o') and blue
(`*') points represent the phase mixing time estimated for
$eB_0/(mc\omega_p) = 0.45$ and $0.1$, respectively. The solid lines are
the linear least square fits to the respective plots corresponding to the
scaling $\delta^{-0.81}$ (for $eB_0/(mc\omega_p) = 0.45$) and $\delta^{-0.86}$ (for
$eB_0/(mc\omega_p) = 0.1$) respectively. (b) Linear plots of phase-mixing time
versus $eB_0/(mc\omega_p)$ estimated for $\delta  = 0.001$.}

\label{fig:figure3}

\end{figure}

{The inspection of the damped modes provides an estimation of the
phase-mixing time, $t_{\rm{mix}}$, according to the definition mentioned above.
In Fig.  \ref{fig:figure3} (a) we show the logarithmic plots of $\omega_p
t_{\rm{mix}}$ as a function of $\delta$ for the two fixed values of $eB_0/(m c
\omega_{p}) = 0.45$ and $0.1$ (denoted by the red and blue points)
respectively. A linear least square fit to these plots (solid lines) reveals
that $\omega_p t_{\rm{mix}} \propto \delta^{-0.86}$ for $eB_0/(m c \omega_{p}) =
0.45$ and $\propto \delta^{-0.81}$ for $eB_0/(m c \omega_{p}) = 0.1$. So, the
scaling of the phase-mixing time as a function of $\delta$ suggested by our
simulations is approximately $\delta^{-0.83}$. We note that the scaling is different
from that estimated for the mixed upper-hybrid mode for which $\omega_p
t_{\rm{mix}} \propto \delta^{-3}$ \cite{maity2013phase}.

{The dependence of the phase-mixing time on $B_0$ is found to be more
complex. Fig. \ref{fig:figure3} (b) shows the linear plot of $\omega_p
t_{\rm{mix}}$ as a function of $eB_0/(m c \omega_{p})$ for the fixed value
$\delta = 0.001$. We find that when $eB_0/(m c \omega_{p}) = 0$, which is the
case of unmagnetized e-p plasma, phase-mixing time is $\sim 200$. For non-zero
$eB_0/(m c \omega_{p})$ in the range of $0$-$0.2$, the phase-mixing time does
not change significantly. Upon further increase of $B_0$ a steep increase of
the phase-mixing time is noticed, so that $\omega_p t_{\rm{mix}} \sim 2500$ at
$eB_0/(m c \omega_{p}) = 0.75$. For the chosen range of $eB_0/(m c \omega_{p})$
in Fig.  \ref{fig:figure3} (b), $\omega_p t_{\rm{mix}}$ is not found to follow
a single  power law. This again is in contrast to the analysis provided for the
mixed-upper hybrid mode \cite{maity2013phase}.}

In the next section we demonstrate the existence of an unique nonlinear
solution 
which does not show any signature of phase mixing 
{in a magnetized e-p plasma}.

\section{Nonlinear Upper-Hybrid Waves }

In this section, we construct a nonlinear solution for a magnetized e-p 
{plasma}
which does not exhibit any phase mixing.  In order to do that we choose the
following initial conditions, 
\begin{eqnarray}
n_e^{(1)}(x,0) = n_{0}\frac{\delta}{2}\cos(kx), 
                 \hspace{0.2cm} n_p^{(1)}(x,0) = -n_{0}\frac{\delta}{2}\cos(kx), \nonumber \\ 
v_{ey}^{(1)}(x,0) = \frac{\delta}{2}\frac{\omega_c}{k} \sin(kx), 
                 \hspace{0.2cm} v_{py}^{(1)}(x,0) =  \frac{\delta}{2}\frac{\omega_c}{k} \sin(kx), \nonumber \\
v_{ex}^{(1)}(x,0) = -\frac{\delta}{2}\frac{\omega_h}{k} \sin(kx), 
                 \hspace{0.2cm} v_{px}^{(1)}(x,0) = \frac{\delta}{2}\frac{\omega_h}{k} \sin(kx). \nonumber \\
\  
\end{eqnarray}
{These}
initial conditions lead to the 
{following}
first order 
{solutions}, 
\begin{eqnarray}
\label{solne11_tw}
   n_{e}^{(1)} &=& -\frac{n_{0}\delta}{2} \cos(kx-\omega_{h}t), \\
\label{solnp11_tw}
   n_{p}^{(1)} &=& \frac{n_{0}\delta}{2} \cos(kx-\omega_{h}t), \\ 
\label{solvex11_tw}
   v_{ex}^{(1)} &=& -\frac{\delta}{2}\frac{\omega_h}{k} \cos(kx-\omega_{h}t), \\
\label{solnvpx11_tw}
   v_{px}^{(1)} &=& \frac{\delta}{2}\frac{\omega_h}{k} \cos(kx-\omega_{h}t), \\
\label{solvey11_tw}
   v_{ey}^{(1)} &=& \frac{\delta}{2}\frac{\omega_c}{k} \sin(kx-\omega_{h}t), \\
\label{sollvpy11_tw}
   v_{py}^{(1)} &=& \frac{\delta}{2}\frac{\omega_c}{k} \sin(kx-\omega_{h}t), \\
\label{solE11_tw}
  E_x^{(1)} &=& \delta \frac{m \omega_{p}^2}{k e} \sin(kx-\omega_{h}t).
\end{eqnarray}
{The above equations correspond to}
a pure traveling wave solution in the first order.  Now in the second order we
choose our initial conditions such that $n_d^{(2)}$ becomes zero. 
{As} 
a result we obtain 
{the}
following relations,
\begin{eqnarray}\label{E2_wave}
n_e^{(2)} = n_p^{(2)}, E_x^{(2)} = 0,  
v_{ex}^{(2)} =  v_{px}^{(2)}, 
v_{ey}^{(2)} = - v_{py}^{(2)}. 
\end{eqnarray}

{Thus, the symmetry recognized in Sects. \ref{sec:2ndorder} and
\ref{sec:3rdorder} [see Eq. \eqref{symmetry1}] is also valid for the present
mode. The symmetry observation for the present mode can also be inferred from
the arguments similar to those presented in Sect. \ref{sec:3rdorder}. Using the
relations in Eq. \eqref{E2_wave} we obtain all the second order quantities from
the following solutions,}
\begin{eqnarray}
\label{solne22_tw}
   n_{e}^{(2)}  &=& -\frac{3 n_{0}\delta^2 \omega_{h}^2}{4(\omega_{c}^2-4\omega_{h}^2)} \cos(2kx-2\omega_{h}t), \\ 
\label{solvex22_tw}
   v_{ex}^{(2)} &=& -\frac{\delta^2}{8} \frac{\omega_{h}}{k}\frac{(\omega_{c}^2+2\omega_{h}^2)}{(\omega_{c}^2-4\omega_{h}^2)}
                    \cos(2kx-2\omega_{h}t), \\
\label{solvey22_tw}
   v_{ey}^{(2)} &=& \frac{3 \delta^2 \omega_{h}^2 \omega_{c}}{8 k (\omega_{c}^2-4\omega_{h}^2)} \sin(2kx-2\omega_{h}t),   
\end{eqnarray}
As
in the first order, the second order solution too exhibit a traveling wave
solution without any sign of {phase-mixing}.
We thus expect that the full {nonlinear} 
solution of the mode will also be oscillatory in space and time {because the ponderomotive 
force, a responsible candidate for phase-mixing, is zero for wave-like solutions}. Thus we
demonstrate that it is possible to have a oscillatory upper-hybrid mode in the
e-p plasma.

{\section{Summary and discussion}}

{We have combined analytical approach and simulations to investigate the
nonlinear evolution of pure upper-hybrid oscillations in a cold magnetized e-p
plasma. In our analytical approach we have used a perturbative analysis where
the relevant physical quantities are expanded in terms of the powers of the
small parameter $\delta$ which corresponds to the maximum amplitude of the
electric field in the system. Different terms in the series are referred to as
the solution of the corresponding order. We have presented analytical solutions
of the relevant quantities up to the third order. Through our solutions we have
demonstrated that in the presence of the magnetic field the pure upper-hybrid
mode does not grow in time until the second order solutions. Our analysis also
show that the slow component of the Lorentz force balances the direct effect of
the ponderomotive forces, therefore we do not see any fast secular terms ($\sim
t$ or $\sim t^2$) in the second order solution. However, self-consistently
generated cyclotron mode along with indirect effect of the ponderomotive forces
introduce {spatial dependency in the frequency of the system } 
which triggers the phase mixing
nonlinearly.} 

{Using the similarities between an electron and a positron we also made a
connection between the symmetry of the dynamical equations to that found in the
non-linear upper-hybrid mode. An interesting consequence of the symmetry is
that the sum of the density fluctuations vanish in the third order, although
third order solutions in general grow with time as $t \cos{\omega_h t}$ where
$\omega_h$ is the upper-hybrid frequency. Through a 1D PIC simulation we have
then confirmed our analytical results and have analyzed the evolution of the
primary mode of oscillations in detail. We have estimated the phase-mixing
time, after which the primary mode decays to {``$1/e$" of initial amplitude,}  
using our simulations.  Our
analysis show that the phase-mixing time $\propto \delta^{-0.83}$ whereas its
dependence on the magnetic field does not follow a simple power law. However,
our analysis clearly demonstrate that the phase mixing of electrostatic
oscillations in a  magnetized e-p plasma is significantly delayed compared to
that in an unmagnetized e-p plasma. In addition, our above findings illustrate
striking contrasts between the pure upper-hybrid oscillations studied in this
paper and the mixed upper-hybrid mode studied in the Ref.
\onlinecite{maity2013phase}.}

{Furthermore, we have presented a generalization of the above mentioned symmetry
identified in the nonlinear solution of the pure upper-hybrid mode to a
traveling wave propagating at the upper-hybrid frequency.  With the help of
this symmetry we have constructed an unique nonlinear solution in a magnetized
e-p plasma which does not show any signature of phase mixing. As a result
the wave continues to propagate at the upper-hybrid frequency without damping.}

In our analysis we have neglected the thermal effects. These effects are 
negligible in many situations of interests 
where {the frequency $\omega >> k v_{th}$ \cite{davidson1972methods}, here 
$v_{th}$ stands for the thermal velocity and $k$ is the wavenumber.}
However, for more general situations the thermal and viscous effects should be
considered. Investigations along these directions are in progress and would be
reported else where.  Examining the role of the relativistic effects on the
upper-hybrid oscillations in an e-p plasma, which has direct applications in the
wake-field acceleration experiments, could also be a possible future direction
of our work. We conclude by noting that our results are relevant for
experiments as in real situations it is natural to excite a single frequency
coherent mode that is not accompanied by any specific zero-frequency mode and
hence the results are important to understand and design probable controlled
experiments.

\appendix*

\section{Third order solutions}
\label{appen}

From the set of Eqs. \eqref{econ5}-\eqref{poi5} we construct a second order partial differential equation 
for $n_d^{(3)}$ similar to the one in Sect. \ref{sec:1storder} for $n_d^{(1)}$ and obtain the 
solution as,  
\begin{widetext}
\begin{eqnarray}
n_d^{(3)} &=& \frac{\delta^3 n_0 \cos{k x} }{64 \omega_c^3 \left[\omega_c^2 - 4 \omega_h^2\right)^2} 
              \left\{\vphantom{\frac{1}{2}}  
                          64  \omega_c^4 \omega_h^3 \left[ 2 - 3 \cos{(2 k x)}\right] \cos{(\omega_h t)}   \sin{(\omega_c t)}
                       - 128  \omega_c^2 \omega_h^5 \left[2 - 3  \cos{(2 k x)}\right]  \cos{(\omega_h t)} \sin{(\omega_c t)} \right. \nonumber \\ 
          &&  \left.  +   64  \omega_h^7            \left[2 - 3  \cos{(2 k x)}\right]  \cos{(\omega_h t)} \sin{(\omega_c t)} 
                      +    4  \omega_c^3 \omega_h^4 
                      \left[ \vphantom{\frac{1}{2}}  
                                       34 \sin{(\omega_c t - \omega_h t)}
                                      -93 \sin{(\omega_h t)}  
                                     +138 \cos{(2 k x)} \sin{(\omega_h t)} \right. \right. \nonumber \\ 
          &&  \left.  \left.  +132 \cos{(2 k x)} \cos{(\omega_c t)} \sin{(\omega_h t)} 
                                      -45  \cos{(2 k x)} \cos{(2\omega_h t)} \sin{(\omega_h t)}
                                      +13 \sin{(3\omega_h t)}
                                      -34 \sin{(\omega_c t + \omega_h t)} \right. \right. \nonumber \\
          &&  \left.  \left. 
                                      +18 \omega_h \left[ 2 - 3 \cos{(2 k x)}\right] t\cos{(\omega_h t)}   
                      \vphantom{\frac{1}{2}}\right]
                        - 32 \omega_c \omega_h^6 
                      \left[ \vphantom{\frac{1}{2}}
                                       4 \left[ 3 - 2 \cos{(\omega_h t)}\right] \sin{(\omega_h t)} 
                                     + 3 \left[ 7 + 5 \cos{(\omega_h t)}\right] \sin{(\omega_h t)} \cos{(2 k x)} \right. \right. \nonumber \\
          &&  \left.  \left.
                                     + 2 \omega_h \left[ 2 - 3 \cos{(2 k x)}\right] t\cos{(\omega_h t)} 
                      \vphantom{\frac{1}{2}}\right]
                      + \omega_c^7 
                      \left[\vphantom{\frac{1}{2}}
                                     - \left[4 - 9 \cos{(2 k x)}\right]\left[ 2 + \cos{(2 \omega_h t)}\right] \sin{(\omega_h t)}
                                                                                                                 \right. \right. \nonumber \\
          &&  \left.  \left.         - 2 \omega_h \left[2 - 3 \cos{(2 k x)}\right] t \cos{(\omega_h t)} 
                      \vphantom{\frac{1}{2}}\right]
                      + \omega_c^5 \omega_h^2 
                      \left[ \vphantom{\frac{1}{2}}
                                     - 8 \sin{(\omega_c t - \omega_h t)} 
                                     + 129 \sin{(\omega_h t)} 
                                     - 222 \cos{(2 k x)} \sin{(\omega_h t)} \right. \right. \nonumber \\ 
          &&  \left.  \left.         - 48 \cos{(2 k x)} \cos{(\omega_h t)} \sin{(\omega_h t)} 
                                     + 9 \cos{(2 k x)} \cos{(2 \omega_h t)} \sin{(\omega_h t)}   
                                     - 5 \sin{(3\omega_h t)}
                                     + 8 \sin{(\omega_c t + \omega_h t)}  \right. \right. \nonumber \\ 
          &&  \left.  \left.         - 6 \omega_h \left[2 - 3 \cos{(2 k x)}\right] t \cos{(\omega_h t)}  
                      \vphantom{\frac{1}{2}}\right] 
              \vphantom{\frac{1}{2}}\right\} \\
E_x^{(3)} &=& \frac{\delta^3 (m\omega_p^2/e) \sin{k x} }{128 k \omega_c^3 \left[\omega_c^2 - 4 \omega_h^2\right)^2} 
              \left\{ \vphantom{\frac{1}{2}}
                      \omega_c^7
                      \left[ \vphantom{\frac{1}{2}}
                                       3 \left[2 + 3 \cos{(2 k x)}\right] \sin{(\omega_h t)}
                                     +   \left[2 + 3 \cos{(2 k x)}\right] \sin{(3 \omega_h t)}
                      \vphantom{\frac{1}{2}}\right]  \right. \nonumber \\
          &&  \left.  + \omega_c^7 \omega_h 4 \cos{(2 k x)} \cos{(2 \omega_h t)} 
                      + 2 \omega_c^5 \omega_h^2
                      \left[ \vphantom{\frac{1}{2}}
                                        8 \sin{(\omega_c t - \omega_h t)}
                                     - 22 \sin{(\omega_h t)} 
                                     - 74 \cos{(2 k x)} \sin{(\omega_h t)}      \right. \right. \nonumber \\  
          &&  \left.  \left.         - 16 \cos{(\omega_c t)} \sin{(\omega_h t)}  
                                     +  3 \cos{(3 \omega_h t)} \sin{(\omega_h t)}
                                     -  2 \sin{(3 \omega_h t)} 
                                     -  8 \sin{(\omega_c t + \omega_h t)}
                      \vphantom{\frac{1}{2}}\right]                                     \right. \nonumber \\ 
          &&  \left.  + 4 \omega_c^4 \omega_h^3
                      \left[ \vphantom{\frac{1}{2}}
                                     - 32 \cos{(2 k x)} \cos{(\omega_h t)} \sin{(\omega_h t)}
                                     +  3 \omega_c \cos{(2 k x)} t \cos{(\omega_h t)} 
                      \vphantom{\frac{1}{2}}\right] 
                      + 8 \omega_c^3 \omega_h^4
                      \left[ \vphantom{\frac{1}{2}}
                                     - 10 \sin{(\omega_c t - \omega_h t)}              \right. \right. \nonumber \\ 
          &&  \left.  \left.         + 14 \sin{(\omega_h t)} 
                                     + 46 \cos{(2 k x)} \sin{(\omega_h t)}           
                                     + 44 \cos{(2 k x)} \cos{(\omega_h t)} \sin{(\omega_h t)}
                                     - 15 \cos{(2 k x)} \cos{(2 \omega_h t)} \sin{(\omega_h t)}
                                                                                        \right. \right. \nonumber \\
          &&  \left.  \left.         -  2 \sin{(3\omega_h t)}
                                     + 10 \sin{(\omega_c t + \omega_h t)}   
                      \vphantom{\frac{1}{2}}\right]  
                      + 16 \omega_c^2 \omega_h^5
                      \left[ \vphantom{\frac{1}{2}}
                                       16 \cos{(2 k x)} \cos{(\omega_h t)} \sin{(\omega_c t)}
                                     -  9 \omega_c \cos{(2 k x)} t \cos{(\omega_h t)}
                      \vphantom{\frac{1}{2}}\right]                                          \right. \nonumber \\
          &&  \left.  - 64 \omega_c \omega_h^6
                      \left[ \vphantom{\frac{1}{2}}
                                        2 \sin{(\omega_h t)} 
                                     +  7 \cos{(2 k x)}\sin{(\omega_h t)}
                                     +  2 \cos{(\omega_c t)} \sin{(\omega_h t)}
                                     +  5 \cos{(2 k x)} \cos{(\omega_c t)} \sin{(\omega_h t)}
                      \vphantom{\frac{1}{2}}\right]                                     \right. \nonumber \\
          &&  \left.  - 128 \omega_h^7
                      \left[ \vphantom{\frac{1}{2}}
                                          \cos{(2 k x)} \cos{(\omega_h t)} \sin{(\omega_c t)}
                                    -  \omega_c \cos{(2 k x)} t \cos{(\omega_h t)}
                      \vphantom{\frac{1}{2}}\right]
              \vphantom{\frac{1}{2}}\right\} 
\end{eqnarray}
\end{widetext}

%

\end{document}